\documentclass[a4paper,twocolumn]{article}
\input{epsf}
\input{psbox}
\newcommand{\nc}{\newcommand}
\nc {\ber} {\begin{eqnarray}}
\nc {\eer} {\end{eqnarray}}
\nc {\beq} {\begin{equation}}
\nc {\eeq} {\end{equation}}
\newcommand{\tl} {\tau_{L}}
\newcommand{\tr} {\tau_{R}}
\newcommand{\dl} {\delta_{L}}
\newcommand{\dr} {\delta_{R}}
\newcommand{\ml} {\left( \begin{array}{cc}
\tl & 1 \\
-\dl & 0
\end{array} \right)}
\newcommand{\mr} {\left( \begin{array}{cc}
\tr & 1 \\
-\dr & 0
\end{array} \right)}
\newcommand{\gr}{\Gamma_\rho}
\newcommand{\ux} {\left( \begin{array}{c} 1\\0 \end{array} \right)}

\newcommand{\xy} {\left( \begin{array}{c} x\\y \end{array} \right)}           

\newcommand{\gu} {G_{\mu}}
\nc {\uml} {{\bf U}_L}
\nc {\umr} {{\bf U}_R}
\nc {\sml} {{\bf S}_L}
\nc {\smr} {{\bf S}_R}

\begin{document}

\bibliographystyle{IEEE}
\large
\title{Robust Chaos}
\author{Soumitro Banerjee $^{1,*}$, James A. Yorke $^{2,\dag}$ and Celso
Grebogi $^{2,\ddag} $\\ $^1$ {\em \normalsize Department of Electrical
Engineering, Indian Institute of Technology, Kharagpur--721302, India} \\ $^2$
{\em \normalsize Institute of Physical Science \& Technology, University of
Maryland, College Park, MD 20742, USA}} 
\date{\normalsize \begin{quote} Practical applications of chaos require the
chaotic orbit to be robust, defined by the absence of periodic windows and
coexisting attractors in some neighborhood of the parameter space. We show
that robust chaos can occur in piecewise smooth systems and obtain the
conditions of its occurrence. We illustrate this phenomenon with a practical
example from electrical engineering. \end{quote}}
\maketitle

\normalsize

It has been proposed to make practical use of chaos in communication
\cite{Hayes93}, in enhancing mixing in chemical processes \cite{Ottino} and in
spreading the spectrum of switch-mode power suppies to avoid electromagnetic
interference \cite{Isabelle-Verghese,jhbd1}. In such applications it will be
necessary to obtain reliable operation in the chaotic mode. 

It is known that for most smooth chaotic systems (take the logistic map
\cite{Gra} for example), there is a dense set of periodic windows for any
range of parameter values. Therefore in practical systems working in chaotic
mode, slight inadvertent 
fluctuation of a parameter may take the system out of chaos. The question is,
how to guarantee that there is no periodic window for a given range of
parameter values and the maximal Lyapunov exponent remains positive throughout
the range? In this Letter, we address this problem.

We say a chaotic attractor is {\em robust} if, for its parameter values there
exists a neighborhood in the parameter space with no periodic attractor and
the chaotic attractor is unique in that neighborhood. It is known that robust
chaos cannot occur in smooth systems. In this Letter we show that such
situations can occur in piecewise smooth maps and obtain the conditions of
existence of robust chaos.

\begin{figure}[tbh]
\begin{center}
\mbox{\psboxto(2.2in;0pt){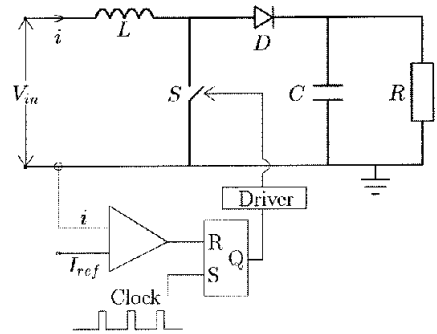}}
\end{center}
\caption{The current mode controlled boost converter}
\label{boostckt}
\end{figure}

We first give a practical example from electrical engineering to demonstrate
robust chaos. The circuit shown in Fig.\ref{boostckt} is known as the boost
converter.  It consists of a controlled switch $S$, an uncontrolled switch
$D$, an inductor $L$, a capacitor $C$ and a load resistor $R$.  When the
controlled switch is turned on, the current in the inductor increases and
energy is stored in it. When the controlled switch is turned off, the stored
energy in the inductor drops and the polarity of the inductor voltage changes
so that it adds to the input voltage. The voltage across the inductor and the
input voltage together ``boosts'' the output voltage to a value higher than
the input voltage. Such circuits are widely used in regulated dc switch-mode power supplies.

Regulation of the output current is achieved by controlling the switching by
current feedback --- known as ``current-mode control''. In this control logic,
the switch is turned on by clock pulses that are spaced $T$ seconds apart.
When the switch is closed, the inductor current increases till it reaches the
specified reference value $I_{ref}$. The switch opens when $i\!=\!I_{ref}$.
Any clock pulse arriving during the {\em on} period is ignored. Once the
switch has opened, the next clock pulse causes it to close.

We obtain a discrete-time model by observing the state variables at every
clock instant. There are two ways in which a state can evolve from one clock
instant to the next. If the on-time $T_{on}\!=\!L(I_{ref}-i_n)/V_{in}$ is less
than $T$, the evolution between observation instants includes one {\em on}
period and one {\em off} period.  Since the clock period typically is much smaller than
the characteristic time of the $LCR$ circuit, we assume the waveforms to be
linear between clock instants. By neglecting the higher order Taylor terms,
the two dimensional map for $T_{on}\!<\!T$ is derived as:
\begin{eqnarray*}
i_{n+1}&=& \!I_{ref} + \frac{1}{L}\left( V_{in}-v_n+\frac{v_nT_{on}}{CR}
\right)\left(T-T_{on} \right)\\
v_{n+1}&=&\!v_n\!-\!\frac{v_nT_{on}}{CR}\!+\!\left(
\frac{I_{ref}}{C}\!-\!\frac{v_n}{CR}\!+\!\frac{v_nT_{on}}{C^2R^2}\right)\left
(T\!-\!T_{on}\right)
\end{eqnarray*}

On the other hand, if the clock pulse arrives while $i\!<\!I_{ref}$, the
switch remains {\em on} between the observation instants. If $T_{on}\ge T$,
then the map takes the form
\begin{eqnarray*}
i_{n+1}&=&i_n+\frac{V_{in}}{L}T\\ v_{n+1}&=& v_n-\frac{v_n}{CR}T
\end{eqnarray*}

The borderline between the two cases is given by the case where the current reaches $I_{ref}$ exactly at
the arrival of the next clock pulse, i.e., $I_{border}\!=\!I_{ref}-V_{in}T/L$.
The resulting map, therefore, is piecewise smooth.

\begin{figure}[tbh]
\begin{center} \mbox{\psboxto(2.2in;0pt) {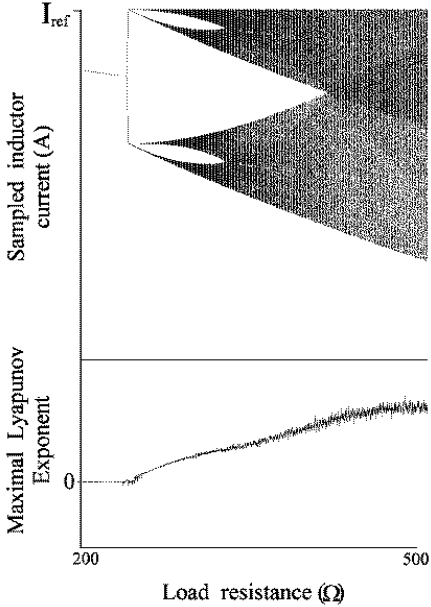}} \end{center}
\caption{The bifurcation diagram and the Lyapunov spectrum of the boost converter. The parameter values are: $c=220\mu F$, $I_{ref}=0.5A$, $V_{in}=30V$, $T=400\mu \! s$, $L=0.1H$}
\label{boostbif}
\end{figure}

The bifurcation diagram and the Lyapunov spectrum of the boost converter is
presented in Fig.\ref{boostbif}. It may be noted that there is no periodic
window or coexisting attractor in the parameter range $R\!=\![241,500]\Omega
$. The chaotic attractor therefore satisfies the conditions of robustness.

We now obtain the general conditions of occurrence of robust chaos.  Let
$f(\hat{x},\hat{y};\rho)$ be a two-dimensional piecewise smooth map which
depends on a single parameter $\rho$. Let $\gr$, given by
$\hat{x}=h(\hat{y},\rho)$ denote a smooth curve that divides the phase plane
into two regions $R_a$ and $R_b$.  The map is given by 
\begin{equation} f(\hat{x},\hat{y};\rho) = \left\{ \begin{array}{ll}
f_1(\hat{x},\hat{y};\rho) \;\; \mbox{for}\;\; \hat{x},\hat{y} \in R_a ,\\
f_2(\hat{x},\hat{y};\rho) \;\; \mbox{for} \;\; \hat{x},\hat{y} \in R_b
\end{array} \right. \label{psm}
\end{equation}
It is assumed that the functions $f_1$ and $f_2$ are both continuous and have
continuous derivatives. The map $f$ is continuous but its derivative is
discontinuous at the line $\gr$, called the ``border''. It is further assumed
that the one-sided partial derivatives at the border are finite. We study the
bifurcations of this system as the parameter $\rho$ is varied.

If a bifurcation occurs when the fixed point of the map is in one of the
smooth regions $R_a$ or $R_b$, it will be one of the ``standard'' types,
namely period doubling, saddle-node or Hopf bifurcation. But if the
bifurcation occurs when the fixed point is {\em on} the border, there is a
discontinuous change in the elements of the Jacobian matrix as $\rho$ is
varied. A rich variety of bifurcations have been reported
\cite{Nusse92,Nusse94,Nusse95} in this situation, which have been called
border collision bifurcation. We show that under certain conditions border
collision bifurcation results in robust chaos. 

It has been shown \cite{Nusse92} that by a change of coordinates, any
piecewise smooth map can be reduced to the normal form (\ref{normal}) in some
small neighborhood of the fixed point undergoing border collision bifurcation.
\begin{eqnarray}
\gu = \left\{ \begin{array}{ll} \ml \xy + \mu \ux, & \mbox{for} \; x \leq 0,
\\
\mr \xy + \mu \ux, & \mbox{for} \; x > 0, \end{array} \right. \label{normal}
\end{eqnarray}
\noindent where $x$ and $y$ are the new coordinates for which the border is
along the line $x\!=\!0$, dividing the phase space into two halves $L$ and
$R$. $\mu$ is the new parameter, which is obtained by scaling $\rho$. $\tl$
and $\dl$ are the trace and determinant of the Jacobian matrix in side $L$.
$\tr$ and $\dr$ are the corresponding values in side $R$. Since the trace and
determinant are invariant under change of coordinates, these quantities are
the same as in map $f$, calculated in the neighborhood of the point where
border collision occurs. As the parameter $\mu$ is varied through zero, local
bifurcations depend only on 
the values of $\tl, \dl, \tr$, and $\dr$ appearing in (\ref{normal}) and
therefore it suffices to study the bifurcations in the normal form
(\ref{normal}) in exploring the border collision bifurcations in the piecewise
smooth map (\ref{psm}).

The fixed points of the system in the two sides are given by
\begin{eqnarray*}
L^* &=& \left( \frac{\mu}{1-\tl+\dl},\frac{-\dl \mu}{1-\tl+\dl} \right) \\ R^*
&=& \left( \frac{\mu}{1-\tr+\dr},\frac{-\dr \mu}{1-\tr+\dr} \right)
\end{eqnarray*}
and their stability is determined by the eigenvalues
$\lambda_{1,2}\!=\!\frac{1}{2}\left( \tau \pm \sqrt{\tau^2-4\delta} \right) $.

It may be noted that if \beq \tl \!>\!(1+\dl) \;\;\; \mbox{and} \;\;\;\tr
\!<\!(1+\dr) \label{pair} \eeq then there is no fixed point for $\mu\!<\!0$
and there are two fixed points, one each in $L$ and $R$, for $\mu\!>\!0$. The
two fixed points are born on the border at $\mu\!=\!0$. We call this a {\em
border collision pair} bifurcation. An analogous situation occurs if $\tl
\!<\!(1+\dl)$ and $\tr \!>\!(1+\dr)$ as $\mu$ is reduced through zero. Due to
symmetry of the two cases, we consider only the parameter region (\ref{pair}).

If $(1+\dr)\!>\tr\!>-(1+\dr)$ then for $\mu\!>\!0$, the fixed point in $L$ is
a regular saddle and the one in $R$ is an attractor. This is like a
saddle-node bifurcation occurring on the border. Since this region in the
parameter space always has a periodic attractor for $\mu\!>\!0$, we exclude
this region from our analysis when looking for chaotic behavior. The condition
$\tr\!=\!-(1+\dr)$, results in a nongeneric situation where all points on the
line joining the points $(\frac{\mu}{1+\dr},0)$ and $(0,-\frac{\dr
\mu}{1+\dr})$ are fixed points of the second iterate. We therefore concentrate
on the parameter space region 
\beq \tl \!>\!(1+\dl) \;\;\; \mbox{and} \;\;\;\tr \!<\!-(1+\dr) \label{pair2}
\eeq and investigate the property of the attractor for $\mu\!>\!0$. We first
consider the case $1\!>\!\dl\!\geq\!0$ and $1\!>\!\dr\!\geq\!0$. 

For (\ref{pair2}), $L^*$ is a regular saddle and $R^*$ is a flip saddle. Let
$\uml$ and $\sml$ be the unstable and stable manifold of $L^*$ and $\umr$ and
$\smr$ be the unstable and stable manifold of $R^*$ respectively. For
(\ref{normal}), all intersections of the unstable manifolds with $x\!=\!0$ map
to the line $y\!=\!0$. Since one linear map changes to another linear map
across the $x\!=\!0$ line, $\uml$ and $\umr$ experience folds along the
x-axis. And all images of fold points will be fold points. By a similar
argument we conclude that $\sml$ and $\smr$ fold along the y-axis, and all
pre-images of fold points are fold points.

Let $\lambda_{1L}$, $\lambda_{2L}$ be the eigenvalues in side $L$ and
$\lambda_{1R}$, $\lambda_{2R}$ be the eigenvalues at side $R$. For condition
(\ref{pair2}), $\lambda_{1L}\!>\!\lambda_{2L}\!>\!0$ and
$0\!>\!\lambda_{1R}\!>\!\lambda_{2R}$.  The stable eigenvector at $R^*$ has a
slope $m_1\!=\!(-\dr/\lambda_{1R})$ and the unstable eigenvector has a slope
$m_2\!=\!(-\dr/\lambda_{2R})$. Since points on an eigenvector map to points on
the same eigenvector and since points on the y-axis map to the x-axis, we
conclude that points of $\umr$ to the left of y-axis map to points above
x-axis. From this we find that $\umr$ has an angle $m_3\!=\!\frac{\dl
\lambda_{2R}}{\dr-\tl \lambda_{2R}}$ after the first fold. Under condition
(\ref{pair2}) we have $m_1\!>\!m_2\!>\!0$ and $m_3\!<\!0$. Therefore there
must be a transverse homoclinic intersection in $R$. This implies an infinity
of homoclinic intersections and the existence of a chaotic orbit. 

\begin{figure}[tbh]
\begin{center}
\mbox{\psboxto(2.2in;0pt){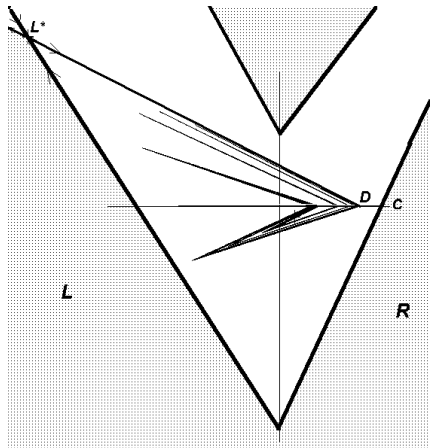}}
\end{center} 
\caption{The stable and unstable manifolds of $L^*$ for $\tl\!=\!1.7$, $\dl\!=\!0.5$, $\tr\!=\!-1.7$, $\dr\!=\!0.5$. $R^*$ is marked by the small cross inside the attractor.}
\label{trap}
\end{figure}

We now investigate the stability of this orbit. The basin boundary is formed
by $\sml$. $\sml$ folds at the y-axis and intersects the x-axis at point $C$.
The portion of $\uml$ to the left of $L^*$ goes to infinity and the portion to
the right of $L^*$ leads to the chaotic orbit.  $\uml$ meets the x-axis at
point $D$, and then undergoes repeated foldings leading to an intricately
folded compact structure as shown in Fig.\ref{trap}.

The unstable eigenvector at $L^*$ has a negative slope given by
$(-\dl/\lambda_{1L})$. Therefore it must have a heteroclinic intersection with
$\smr$. Since both $\uml$ and $\umr$ have transverse intersections with
$\smr$, by the Lambda Lemma \cite{yorkebook} we conclude that for each point
$q$ on $\umr$ and for each $\epsilon$-neighborhood $N_\epsilon (q)$, there
exist points of $\uml$ in $N_\epsilon (q)$. Since $\uml$ comes arbitrarily
close to $\umr$, the attractor must span $\uml$ in one side of the
heteroclinic point.

Since all initial conditions in $L$ converge on $\uml$ and all initial
conditions in $R$ converge on $\umr$, and since there are points of $\uml$ in
every neighborhood of $\umr$, we conclude that the attractor is unique. This
chaotic attractor can not be destroyed by small changes in the parameters.
Since small changes in the parameters can only cause small changes in the
Lyapunov exponents, where the chaotic attractor is stable, it is also robust.

It is clear from the above geometrical structure that no point of the
attractor can be to the right of point $D$.  If $D$ lies towards the left of
$C$, the chaotic orbit is stable. If $D$ falls outside the basin of
attraction, it is an unstable chaotic orbit or chaotic saddle. From this, the
condition of stability of the chaotic attractor is obtained as 
\beq
\dl \tr \lambda_{1L} - \dr \lambda_{1L} \lambda_{2L} + \dr \lambda_{2L} - \dl
\tr + \tl\dl - \dl^2 - \lambda_{2L}\dl >0 \label{cond} \eeq 
If $\dl\!=\!\dr\!=\!\delta$ this condition reduces to
$\tr\lambda_{1L}-\lambda_{1L}\lambda_{2L}+\tl-\tr-\delta\!>\!0 $.

The robust chaotic orbit continues to exist as $\tl$ is reduced below
$(1\!+\!\dl)$. With $\tl$ slightly below $(1\!+\!\dl)$, there is no fixed
point in $L$ for $\mu\!>\!0$ but the invariant manifolds suffer only slight
change. The invariant manifold of $L$ associated with $\lambda_{1L}$ still
forms the attractor. The invariant manifolds in $L$, however, cease to exist
for $\tl \!<\!2\sqrt{\dl}$ since the eigenvalues become complex. As $\tl$ is
reduced below $2\sqrt{\dl}$ there is a sudden reduction in the size of the
attractor as it spans only $\umr$. So long as $\uml$ exists, multiple
attractors can not exist and therefore if the main attractor is chaotic, it is
also robust.

\begin{figure}[tbh]
\begin{center}
\mbox{\psboxto(2.2in;0pt){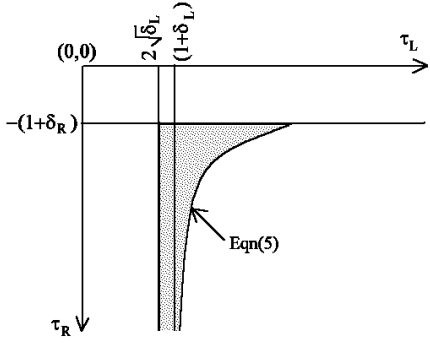}}
\end{center} 
\caption{Schematic diagram of the parameter space region of the normal form (\ref{normal}) where robust chaos is observed for $1\!>\!\dl\!>\!0$, $1\!>\!\dr\!>\!0$ and $\mu\!>\!0$.}
\label{para}
\end{figure}

Therefore we see that for $1\!>\!\dl\!>\!0$, $1\!>\!\dr \!>\!0$, the normal
form (\ref{normal}) exhibits robust chaos in a portion of parameter space
bounded by the conditions $\tr\!=\!-(1+\dr)$, $\tl\!>\! 2\sqrt{\dl}$ and
(\ref{cond}), as shown in Fig.\ref{para}. There is a symmetric region of the
parameter space with the roles of $R$ and $L$ interchanged, where the same
phenomena are observed for $\mu\!<\!0$.

At very low values of the determinant, i.e., when the system is very close to
being one-dimensional, the main attractor may not remain chaotic even for
$\tl\!>\!2\sqrt{\dl}$, as periodic orbits become stable. The conditions for
emergence of periodic windows for the one-dimensional case have been derived
in \cite{Nusse95,Maistrenko}. Therefore for one dimensional systems, the
parameter range for robust chaos is bounded by $\tr\!=\!1$,
$\tr>-\frac{\tl}{\tl-1}$, and the lower limit of $\tl$ is given by the
conditions of existence of various periodic windows. Here the $\tau$'s are to
be interpreted as the slopes of the piecewise linear function in the two
halves separated by $x\!=\!0$.  On the other hand, if the determinants in the
two sides are unity, the region in the $\tl\!-\!\tr$ space for robust chaos
shrinks to zero area.

The cases with negative determinant are investigated following the same
method. For the sake of brevity, we present the results without much
explanation.

For $-1\!<\!\dr\!<\!0$, we have $1\!>\!\lambda_{1R}\!>\!0$,
$\lambda_{2R}\!<\!-1$, and $R^*$ is located above the x-axis. A positive value
of $\lambda_{1R}$ implies that $\uml$ converges on $\umr$ from one side. If
\beq \frac{\lambda_{1L}-1}{\tl-1-\dl}>\frac{\lambda_{2R}-1}{\tr-1-\dr}
\label{whichpoint} \eeq 
then the intersection of $\uml$ with the x-axis remains the rightmost point of
the attractor and (\ref{cond}) still gives the parameter range for boundary
crisis. But if (\ref{whichpoint}) is not satisfied, the intersection of $\umr$
with the x-axis becomes the rightmost point of the attractor, and the
condition of existence of the chaotic attractor changes to
\beq
{\frac { \lambda_{2R}-1}{\tau_{{R}}-1-\delta_{{R}}}}<{\frac {
\delta_{{L}}\left (\tau_{{L}}-\delta_{{L}}- \lambda_{2L}\right )
}{\left (\tau_{{L}}-1-\delta_{{L}}\right )\left (\delta_{{R}}
\lambda_{2L}-\delta_{{L}}\tau_{{R}}\right )}}
\eeq

For $\dl\!<\!0$ and $\dr\!<\!0$, $L^*$ is below the x-axis and the same logic
as above applies. But if $\dl\!<\!0$ and $\dr\!>\!0$, the stable manifold of
$R^*$ has a negative eigenvalue and hence $\uml$ does not approach $\umr$
from one side. Therefore, if (\ref{whichpoint}) is not satisfied, there is no
analytic condition for boundary crisis --- it has to be determined
numerically.
For $\dl\!<\!0$, the invariant manifolds $\uml$ and $\sml$ always exist as the
eigenvalues are real for all $\tl$. Therefore multiple attractors
can not exist for $\dl \!<\!0$.

Since (\ref{normal}) is a normal form of the piecewise smooth map (\ref{psm}),
it is expected that robust chaos would be observable in many piecewise smooth
maps in the neighborhood of border collision bifurcations, provided that there
are no more than one period-1 fixed point in $R_a$ and $R_b$, there
exist homoclinic as well as heteroclinic intersections of the invariant
manifolds associated with these fixed points, and the
trace and determinant at the two sides of the borderline satisfy the above
conditions. The example of the boost converter is a case in point.

A major conclusion of this Letter is that one should use piecewise smooth
systems in applications that require reliable operation under chaos.

\vspace{0.2in}
\noindent $^*$ Electronic address: soumitro@ee.iitkgp.ernet.in\\
$^\dag$ Electronic address: yorke@ipst.umd.edu\\ $^\ddag$ Also with the
Institute for Plasma Research and Department of Mathematics, University of
Maryland, College Park, USA. Electronic address: grebogi@chaos.umd.edu

\end{document}